\documentclass[reprint,
 amsmath,amssymb,
 aps,
]{revtex4-2}

\usepackage{graphicx}
\usepackage{dcolumn}
\usepackage{bm}

\usepackage{physics}
\usepackage{tensor}
\usepackage{xcolor}
\usepackage{babel}
\usepackage{float}
\usepackage{subcaption}

\begin{document}

\preprint{APS/123-QED}

\title{Cosmological Particle Creation Using an Equal-Time Wigner Formalism}

\author{Philip Semr\'{e}n}
  \email{philip.semren@umu.se}

\affiliation{%
Department of Physics, Ume{\aa} University, Sweden\\
}%

\date{\today}

\begin{abstract}

It is well known that the expansion of the universe can create particles. However, due to ambiguities when defining particles during the expansion, there are still debates about how to choose vacuum and particle states. To clarify how particles are produced in an expanding universe,  we study the creation of real scalar particles in flat FLRW spacetimes by using a recently developed equal-time Wigner formalism. By comparing this quantum kinetic formalism with the standard Bogoliubov approach, we make a natural definition of a particle number in terms of kinetic phase-space functions, which we then compare with common adiabatic particle numbers. With inspiration from flat spacetime QED, we perform numerical calculations and discuss the interpretation of the particle numbers in terms of a hypothetical switch-off in the expansion rate. Finally, we consider how this interpretation is affected by regularization.   
\end{abstract}

\maketitle


\section{\label{sec:Introduction} Introduction}

As the early universe expands, particles are produced by the expansion. This process is one of few mechanisms that could produce particle dark matter, and is generally described using the framework of quantum field theory in curved spacetime. However, in this framework there are ambiguities that prevent us from uniquely defining what a particle is during the expansion. This leads to similar ambiguities in the produced number of particles, which has sparked an unresolved debate about whether the vacuum state in de Sitter spacetime is stable or not \cite{ford_cosmological_2021}. To settle the discussions, the competing particle definitions would have to be interpreted in terms of physical particles. 

Clarifying what is meant by a physical particle in an expanding spacetime is, however, not an easy task due to the aforementioned ambiguities. There is therefore a widespread notion that the fundamental quantities should be the quantum fields and their expectation values rather than particles, and when talking about particles they should be given an operational definition, for instance by setting up Unruh-DeWitt detectors. However, even without specifying any detector details, discussions about particle definitions under non-stationary conditions can still provide some valuable and unambiguous information when interpreted in terms of asymptotic stationary regions, where the vacuum ambiguities vanish. In the context of flat-spacetime QED, particle numbers at intermediate times have been interpreted in terms of asymptotic particle numbers obtained after the time-dependent external agent (the electric field) has been switched off \cite{ilderton_physics_2022}. Switching off the electric field has also been used to give the intermediate particle numbers an operational meaning \cite{alvarez-dominguez_operational_2023}. 

Moving focus away from QED, we consider similar interpretations of particle numbers in an expanding universe. In this cosmological context, where the electric field is replaced by the expansion rate of spacetime, the observer has less agency over the external agent. Although the operational interpretation hence loses some of its strength, it still provides an intuitive and unambiguous meaning to the quantities appearing in the chosen formalism.   

The formalism that we will use is based on a recently developed quantum kinetic approach for curved spacetimes \cite{friedrich_kinetic_2018}. This formalism revolves around a set of phase-space functions that serve as the quantum counterparts to the classical distribution function, and determine the energy momentum tensor on integration over the momenta \footnote{Note that, although they play a similar role, the phase-space functions used in this quantum kinetic approach are generated by a Wigner transform, and thus have properties that set them apart from classical distribution functions. For instance, the Wigner functions can be negative. Despite such differences, which are less significant in the spatially homogeneous case, we will occasionally refer to the Wigner functions as distribution functions.}. Thus, the equations are written using quantities that are close to physical interpretation. As a result, we will show that the quantum kinetic approach immediately leads to a natural particle definition, which we interpret in terms of a hypothetical switch-off in the expansion rate. 

Our motivation for using a quantum kinetic approach comes from flat spacetime, where quantum kinetic models have been widely used to incorporate quantum effects when studying plasmas and other inherently statistical systems \cite{brodin_quantum_2022}.  When based on fully relativistic quantum theories, these models can describe fundamental quantum phenomena as well as strong-field effects. Strong-field electron-positron pair production can, for instance, be captured using the Dirac-Heisenberg-Wigner formalism, which uses an equal-time Wigner transform of the Dirac equation in flat spacetime  \cite{bialynicki-birula_phase-space_1991}.   

Although it is quantum field theories in flat spacetimes that serve as the basis for most relativistic quantum kinetic models, analogues applicable to curved spacetimes have also been developed. These have generally used fully covariant approaches \cite{winter_wigner_1985,fonarev_wigner_1994, calzetta_quantum_1988, antonsen_quantum_1997}, where the kinetic transport equations are accompanied by quantum mass-shell constraints. This is also a feature of covariant models in flat spacetime \cite{elze_transport_1986,vasak_quantum_1987}. However, by partially breaking the explicit covariance, the authors of Ref.~\cite{friedrich_kinetic_2018} recently developed an equal-time formalism for real scalar fields in curved spacetime. This approach naturally leads to a set of dynamical equations that are closed and on-shell without extra constraints, similarly to what happens when considering equal-time approaches in flat spacetime \cite{best_phase-space_1993,bialynicki-birula_phase-space_1991}.    

In Ref.~\cite{friedrich_kinetic_2018} the context was to derive equations suitable for describing dark matter, and an emphasis was put on the observation that a certain combination of the equations reproduces the general relativistic collisionless Boltzmann equation in the classical limit. Moving the focus away from the classical limit, we will instead use the framework from Ref.~\cite{friedrich_kinetic_2018} to describe particle creation in Friedmann-Lema\^itre-Robertson-Walker (FLRW) spacetimes due to a dynamical scale factor.       

 Particle numbers and pair creation in cosmology have been studied extensively before (see e.g.~Ref.~\cite{ford_cosmological_2021} for a recent review). The prevailing approach for studying these aspects is to make use of Bogoliubov transformations to relate the incoming and outgoing creation and annihilation operators. Alongside this approach, kinetic phase-space methods have also been used (see e.g.~Refs.~\cite{garbrecht_particle_2004,kainulainen_tachyonic_2023, sobol_schwinger_2020,lysenko_quantum_2023}).  Contributing to the kinetic description, a key result of this paper is the formulation of common particle number definitions in terms of the phase-space functions defined in Ref.~\cite{friedrich_kinetic_2018}. We also interpret the particle numbers in terms of a hypothetical switch-off in the expansion rate, in alignment with results from quantum electrodynamics in flat spacetime \cite{ilderton_physics_2022}. 

The paper will be outlined as follows. First we will give a short summary of the parts of Ref.~\cite{friedrich_kinetic_2018} that are needed for our purposes. That includes the Arnowitt-Deser-Misner (ADM) decomposition and the Wigner transformation of the Klein-Gordon equation, which we then apply to FLRW spacetimes. Then, to make connection with the Bogoliubov approach, we consider expansions of the scalar field in terms of specific known mode functions. That allows us to make a natural particle definition using the phase-space functions of the theory. This definition is then compared, both analytically and numerically, to commonly used adiabatic particle numbers. Finally, we discuss how regularization affects the interpretation of the particle numbers.       

\section{\label{sec:Preliminaries} Preliminaries}
Here we collect some preliminaries that are needed to define the sought phase-space functions and determine their evolution equations. For more details, the reader is referred to Ref.~\cite{friedrich_kinetic_2018}.

\subsection{ADM Decomposition}
To define equal-time Wigner functions, we first need a notion of equal-time surfaces. We therefore assume that we have a globally hyperbolic spacetime that can be foliated in terms of a family of spacelike hypersurfaces $\Sigma_t$, each hypersurface labeled by a certain value of the corresponding level function $t$. To this level function, we define an associated vector field $t^\mu$ through  
\begin{align}
t^\mu \nabla_\mu t = 1, \qq{} t^\mu = Nn^\mu+N^\mu, 
\end{align}
with lapse $N$ and shift $N^\mu$ satisfying $n_\mu N^\mu = 0$. Here $n_\mu$ is assumed to be the normal to the spatial hypersurface, proportional to $(\dd{t})_\mu$ and with norm $n_\mu n^\mu = -1$. Using $t$ as the zero coordinate $x^0$ and latin indices $i,j,k,\ldots$ to run over $1,2,3$,  we can then write 
\begin{align}
t^\mu &= \pmqty{1 & 0}, \qq{} n_\mu = \pmqty{-N & 0}, \\
 N^\mu &= \pmqty{0 & N^i}, \qq{} n^\mu = N^{-1}\pmqty{1 & -N^i}, \\
   g\indices{_\mu_\nu} &= \pmqty{-N^2 +N^iN_i & N_j \\ N_i & \gamma\indices{_i_j}},  \label{eq:ADM_metric}\\
   ~
   g\indices{^\mu^\nu} &=\pmqty{-N^{-2} & N^{-2}N^j \\ N^{-2}N^i & \gamma\indices{^i^j}-N^{-2}N^iN^j}, \\
   ~
   \sqrt{-g} &= N\sqrt{\gamma},
\end{align}
where 
$\gamma\indices{_i_j}$ is the induced metric on the spatial hypersurface, $\gamma\indices{^i^j}$ is its inverse, and $\gamma\indices{_\mu_\nu} = g\indices{_\mu_\nu} + n_\mu n_\nu$ can be seen as a projection tensor projecting onto the spatial hypersurfaces.

Having thus defined a $1+3$ decomposition of spacetime, we consider the dynamics of a real scalar field $\phi$ in this decomposition. 
Using the action for a massive, minimally coupled, scalar field without additional interactions, the equations of motion for $\phi$ and its canonical momentum $\Pi$ can be written as 
\begin{align}
    \partial_t \phi &= \frac{N}{\sqrt{\gamma}}\Pi + N^j\partial_j\phi, \label{eq:phievolution}\\
    \partial_t \Pi &= \partial_j\left(N^j\Pi\right) + \partial_i \left(N\sqrt{\gamma}\gamma\indices{^i^j}\partial_j\phi\right) - \frac{N\sqrt{\gamma}m^2}{\hbar^2}\phi. \label{eq:Pievolution}
\end{align}
These equations are equivalent to the Klein-Gordon equation
\begin{equation}
    \nabla_\mu \nabla^\mu \phi = \frac{m^2}{\hbar^2}\phi,
\end{equation}
on returning to covariant form \cite{friedrich_kinetic_2018}.

Furthermore, decomposing the energy momentum tensor for the scalar field
\begin{align}
     T\indices{_\mu_\nu} = \partial_\mu \phi \partial_\nu \phi - \frac{g\indices{_\mu_\nu}}{2}\left(\partial^\sigma \phi \partial_\sigma \phi + \frac{m^2}{\hbar^2}\phi^2\right),
\end{align}
as 
\begin{align}
    T\indices{^\mu^\nu} = \rho n^\mu n^\nu + P\gamma\indices{^\mu^\nu} + q^\mu n^\nu + n^\mu q^\nu + \pi\indices{^\mu^\nu},
\end{align}
in terms of the projections
\begin{align}
    \rho &= n_\mu n_\nu T\indices{^\mu^\nu},\\
    P &= \frac{1}{3}\gamma\indices{_\mu_\nu} T\indices{^\mu^\nu}, \\ q^\mu &= -\gamma\indices{^\mu_{(\sigma}}n\indices{_{\tau)}} T\indices{^\sigma^\tau}, \\
    \pi\indices{^\mu^\nu} &= \left(\gamma\indices{^\mu_{(\sigma}}\gamma\indices{^\nu_{\tau)}} -\frac{1}{3}\gamma\indices{^\mu^\nu}\gamma\indices{_\sigma_\tau}\right)T\indices{^\sigma^\tau} \equiv T\indices{^{\langle\mu}^{\nu\rangle}},
\end{align}
we find that
\begin{align}
    \rho &= \frac{1}{2}\left(\frac{1}{\gamma}\Pi^2+\frac{m^2}{\hbar^2}\phi^2+\gamma \indices{^i^j}\partial_i \phi \partial_j \phi
\right),
\\
P &=  \frac{1}{2}\left(\frac{1}{\gamma}\Pi^2-\frac{m^2}{\hbar^2}\phi^2-\frac{1}{3}\gamma \indices{^i^j}\partial_i \phi \partial_j \phi
\right), \\
q^i &= -\frac{1}{2}\gamma \indices{^i^j} \left(\frac{\Pi}{\sqrt{\gamma}}\partial_j\phi +  \partial_j\phi\frac{\Pi}{\sqrt{\gamma}} \right), \\
\pi\indices{^i^j} &= \left(\gamma \indices{^i^l}\gamma \indices{^j^k} -  \frac{1}{3}\gamma \indices{^i^j}\gamma \indices{^l^k}\right)\partial_l\phi \partial_k\phi,
\end{align}
where round brackets around a pair of indices denotes a symmetrization. Relative to an observer with 4-velocity $n^\mu$,  $\rho$ is the energy density, $P$ is the isotropic pressure,  $q^\mu$ is the energy flow orthogonal to $n^\mu$, and $\pi\indices{^\mu^\nu}$ is the anisotropic pressure. The latter two satisfy $n_\mu q^\mu =n_\mu \pi\indices{^\mu^\nu} = 0 $, $\pi\indices{^\mu^\nu} = \pi\indices{^\nu^\mu}$, and $\gamma\indices{_\mu_\nu}\pi\indices{^\mu^\nu}=0$, so that $q^0 = \pi\indices{^\mu^0}= \pi\indices{^0^\mu}= \gamma\indices{_i_j}\pi\indices{^i^j}=0$.  
\subsection{\label{sec:Wigner Transformation of the Klein-Gordon Equation} Wigner Transformation of the Klein-Gordon Equation}

In the context of classical kinetic theory, bulk properties of the system, such as its energy momentum tensor, are obtained by taking moments of a phase-space distribution function. To formulate something similar for the scalar field, we can note from the previous section that its energy momentum tensor only involves quadratic monomials of $\phi$, $\Pi$, and $\partial_i \phi$. Hence it could be helpful to perform some sort of Fourier transform of the quadratic monomials, interpreting the conjugate variables as momenta. More specifically, after promoting the fields to operators and imposing canonical commutation relations, we will make use of the equal-time Wigner transform defined in Ref.~\cite{friedrich_kinetic_2018} for two operators $\widehat{X}$ and $\widehat{Y}$ through
\begin{widetext}
\begin{align} \label{eq:WignerTransform}
\widehat{F}_{XY}\left(t,x^i,p_k\right) \equiv \sqrt{\gamma} \int_{T\Sigma_t}\dd[3]{r} \exp(-\frac{i}{\hbar}r^kp_k) \left[\exp(\frac{r^k}{2} \tensor[^{(3)}]{\nabla}{^H_k})\widehat{X}\right]\left[\exp(-\frac{r^k}{2} \tensor[^{(3)}]{\nabla}{^H_k})\widehat{Y}\right],
\end{align}
\end{widetext}
where the integral is performed over the coordinates $r^k$ of the fibre of the tangent bundle $T\Sigma_t$ at $x^i$. In this definition we have introduced the horizontal lift of the covariant derivative on $\Sigma_t$ to $T\Sigma_t$ 
\begin{equation}
    \tensor[^{(3)}]{\nabla}{^H_k} \equiv  \tensor[^{(3)}]{\nabla}{_k} - r^l\tensor[^{(3)}]{\Gamma}{^i_k_l}\pdv{}{r^i},
\end{equation}
where $\tensor[^{(3)}]{\nabla}{_k}$ is the covariant derivative on $\Sigma_t$ and $\tensor[^{(3)}]{\Gamma}{^i_l_k}$ its corresponding Christoffel symbols. It should also be noted that we have neglected a normal ordering procedure for (\ref{eq:WignerTransform}) described in Ref.~\cite{friedrich_kinetic_2018}. Instead of using this procedure, which ensures that finite results are obtained on integrating over the momenta, we postpone the issue of divergent integrals to Sec.~\ref{sec:Regularization} where we discuss regularization in terms of an adiabatic subtraction scheme. 

Choosing $\widehat{X}, \widehat{Y} \in \{\phi, \gamma^{-1/2}\Pi\}$, we get the operators $\widehat{F}_{\phi\phi}, \widehat{F}_{\phi\Pi}, \widehat{F}_{\Pi\phi}, \widehat{F}_{\Pi\Pi}$. Then, following Ref.~\cite{friedrich_kinetic_2018}, we define the phase-space functions \footnote{These definitions share similarities with certain definitions from \cite{best_phase-space_1993} for flat spacetime.}
\begin{align}
    f_1^+ &= \frac{1}{\left(2\pi\hbar\right)^3}\frac{1}{2\hbar}\left[\frac{\omega_p}{\hbar}\expval{\widehat{F}_{\phi\phi}} + \frac{\hbar}{\omega_p}\expval{\widehat{F}_{\Pi\Pi}}\right], \label{eq:f1pdef} \\
    ~
    f_1^- &=  \frac{1}{\left(2\pi\hbar\right)^3}\frac{i}{2\hbar}\left[\expval{\widehat{F}_{\Pi\phi}} -\expval{\widehat{F}_{\phi\Pi}}\right], \label{eq:f1mdef}\\
    ~
    f_2^{\phantom{+}} &=  \frac{1}{\left(2\pi\hbar\right)^3}\frac{1}{2\hbar}\left[\frac{\omega_p}{\hbar}\expval{\widehat{F}_{\phi\phi}} - \frac{\hbar}{\omega_p}\expval{\widehat{F}_{\Pi\Pi}}\right], \label{eq:f2def} \\
    ~
    f_3^{\phantom{+}} &= \frac{1}{\left(2\pi\hbar\right)^3}\frac{1}{2\hbar}\left[\expval{\widehat{F}_{\Pi\phi}} +\expval{\widehat{F}_{\phi\Pi}}\right], \label{eq:f3def}
\end{align}
where
\begin{equation}
    \omega_p = \sqrt{m^2 + \gamma\indices{^i^j}p_ip_j},
\end{equation}
and $\expval{}$ denotes the expectation value with respect to the quantum state of the system.  
These phase-space functions can be used to write the projections of the energy-momentum tensor as
\begin{align}
    \rho = {}& \int \frac{\dd[3]{p}}{\sqrt{\gamma}} \omega_p f_1^+  \notag \\  &+ \frac{\hbar^2}{8}\gamma\indices{^i^j}\tensor[^{(3)}]{\nabla}{_i}\tensor[^{(3)}]{\nabla}{_j}\int \frac{\dd[3]{p}}{\sqrt{\gamma}} \frac{f_1^+ + f_2}{\omega_p}, \\
   ~ 
    P = {}& \frac{1}{3}\gamma\indices{^i^j}\int \frac{\dd[3]{p}}{\sqrt{\gamma}}p_ip_j \frac{f_1^++f_2}{\omega_p}-\int \frac{\dd[3]{p}}{\sqrt{\gamma}} \omega_p f_2  \notag \\
    &-\frac{\hbar^2}{24}\gamma\indices{^i^j}\tensor[^{(3)}]{\nabla}{_i}\tensor[^{(3)}]{\nabla}{_j}\int\frac{\dd[3]{p}}{\sqrt{\gamma}} \frac{f_1^+ + f_2}{\omega_p}, \\
    ~
    q^i ={}& \gamma\indices{^i^j}\int \frac{\dd[3]{p}}{\sqrt{\gamma}}p_j f_1^- -\frac{\hbar}{2}\gamma\indices{^i^j}\tensor[^{(3)}]{\nabla}{_j}\int \frac{\dd[3]{p}}{\sqrt{\gamma}}f_3, \\
    ~
    \pi\indices{^i^j} ={}& \left(\gamma \indices{^i^l}\gamma \indices{^j^k} -  \frac{1}{3}\gamma \indices{^i^j}\gamma \indices{^l^k}\right)\Bigg[ \int \frac{\dd[3]{p}}{\sqrt{\gamma}}p_lp_k \frac{f_1^++f_2}{\omega_p} \notag \\
    &+\frac{\hbar^2}{4}\tensor[^{(3)}]{\nabla}{_l}\tensor[^{(3)}]{\nabla}{_k}\int\frac{\dd[3]{p}}{\sqrt{\gamma}} \frac{f_1^+ + f_2}{\omega_p} \Bigg]. 
\end{align}
Thus, we see that the dynamics of the energy momentum tensor can be fully described using the evolution of the phase-space functions.

The evolution equations for the phase-space functions can in principle be determined by using the definition of the Wigner transform (\ref{eq:WignerTransform}) and the evolution equations for the fields (\ref{eq:phievolution})--(\ref{eq:Pievolution}). However, this procedure is in general quite tedious due to the appearance of terms proportional to the Christoffel symbols in the exponentials. Nevertheless, it has has been done in Ref.~\cite{friedrich_kinetic_2018} to leading order in a spatial gradient expansion in powers of $\hbar$. To avoid the complication with the Christoffel symbols and to simplify our analysis, we will restrict our attention to flat FLRW models, where the three-dimensional Christoffel symbols vanish. With this restriction, there is no need for assumptions involving spatial gradient expansions, allowing us to perform a full quantum treatment of the system.

\section{\label{sec:Flat FLRW} Evolution equations for the flat FLRW case}
The flat FLRW models can be described using the line element
\begin{equation}
    \dd{s}^2 = -N(t)^2\dd{t}^2 +a(t)^2\left(\dd{x}^2 + \dd{y}^2 + \dd{z}^2\right),
\end{equation}
where $N = 1$ when $t$ is chosen as comoving time, and $N = a$ when $t$ is conformal time. In the following applications, $t$ will be comoving time, but we keep $N$ general in this section. Comparing the line element with (\ref{eq:ADM_metric}) it furthermore follows that
\begin{align}
    N^i &= 0,\\
    \gamma\indices{_i_j} &= a^2\delta\indices{_i_j}, \quad \gamma\indices{^i^j} = a^{-2}\delta\indices{^i^j}, \quad \sqrt{\gamma} = a^3,  \\
    \tensor[^{(3)}]{\nabla}{^H_k} &= \partial_k,
\end{align}
and the equations of motion reduce to 
\begin{align}
    \partial_t \phi &= \frac{N}{a^3}\Pi, \\
    \partial_t \Pi &= Na\delta\indices{^i^j}\partial_i\partial_j\phi - \frac{Na^3m^2}{\hbar^2}\phi. 
\end{align}
Using these together with (\ref{eq:WignerTransform})--(\ref{eq:f3def}), we deduce that \footnote{As we have derived them here, these equations do not perfectly coincide with the final equations from Ref.~\cite{friedrich_kinetic_2018} when those are applied to the flat FLRW metric.}

\begin{align}
\dot{f}_1^+ ={}& \left(\frac{\dot{\omega}_p}{\omega_p}+3\mathcal{H}\right)f_2 - \frac{N}{\omega_p}p_j\partial^j f_1^-+\frac{\hbar N}{4\omega_p}\partial_j\partial^j f_3, \\
~
\dot{f}_1^- ={}&-\frac{N}{\omega_p}p_j\partial^j\left(f_1^+ +f_2\right), \\
~
\dot{f}_2^{\phantom{+}} = {}&\left(\frac{\dot{\omega}_p}{\omega_p}+3\mathcal{H}\right)f_1^+ + \frac{N}{\omega_p}p_j\partial^j f_1^- -\frac{\hbar N}{4\omega_p}\partial_j\partial^j f_3 \notag \\ &+\frac{2N\omega_p}{\hbar}f_3,     
 \\
~
\dot{f}_3^{\phantom{+}} ={}& -\frac{2N\omega_p}{\hbar}f_2 + \frac{\hbar N}{4\omega_p}\partial_j\partial^j\left(f_1^+ + f_2\right), 
\end{align}
where $\dot{f} \equiv \partial_t f$, $\mathcal{H}\equiv\dot{a}/a$, $\partial^i f = \gamma\indices{^i^j}\partial_j f$, and
\begin{equation}
    \frac{\dot{\omega}_p}{\omega_p}+3\mathcal{H} = \mathcal{H}\left(2+\frac{m^2}{\omega_p^2}\right).
    \label{eq:omegap_dot_rel}
\end{equation}
At this point, the above evolution equations can be seen as describing a test field propagating on a flat FLRW background. However, if the intention is to couple the phase-space functions to the geometry through the energy-momentum tensor, this tensor, and hence the phase-space functions, have to respect the spacetime symmetries. Although we reserve self-consistent calculations with backreaction for another paper, we will therefore assume, in accordance with the homogeneity and isotropy of the spacetime, that the phase-space functions are spatially homogeneous and that 
\begin{align}
    \int \frac{\dd[3]{p}}{\sqrt{\gamma}}p_i f_1^- &= 0, \label{eq:isotropycondition1}\\
    \left(\gamma \indices{^i^l}\gamma \indices{^j^k} -  \frac{1}{3}\gamma \indices{^i^j}\gamma \indices{^l^k}\right) \int \frac{\dd[3]{p}}{\sqrt{\gamma}}p_lp_k \frac{f_1^++f_2}{\omega_p} &= 0, \label{eq:isotropycondition2}
\end{align}
so that $\rho$ and $P$ are homogeneous while $q^\mu$ and $\pi\indices{^\mu^\nu}$ vanish. With these assumptions, the evolution equations reduce to
\begin{align}
\dot{f}_1^+ &= \mathcal{H}\left(2+\frac{m^2}{\omega_p^2}\right)f_2, \label{eq:f1peqFLRW}\\
~
\dot{f}_1^- &= 0, \\
~
\dot{f}_2^{\phantom{+}} &=  \mathcal{H}\left(2+\frac{m^2}{\omega_p^2}\right)f_1^+  +\frac{2N\omega_p}{\hbar}f_3, \\
~
\dot{f}_3^{\phantom{+}} &= -\frac{2N\omega_p}{\hbar}f_2.  \label{eq:f3eqFLRW}
\end{align}
Since $p_i$ only appears explicitly in these equations through the combination $\gamma\indices{^i^j}p_ip_j =\delta\indices{^i^j}p_ip_j/a^2$ in $\omega_p$, they are inherently isotropic with respect to $p$. Hence, provided that the initial conditions share this isotropy, the conditions (\ref{eq:isotropycondition1})--(\ref{eq:isotropycondition2}) are naturally satisfied, showing their compatibility with the homogeneity assumption.

\section{Distribution Functions From Known Mode Functions}

To solve the evolution equations for the phase-space functions in practice, suitable initial conditions are needed. To determine these conditions, and to make connection with the common Bogoliubov approach for studying particle production in cosmology, it is instructive to look at the phase-space functions in terms of certain known mode functions.

For this purpose, assume that the field $\phi$ is quantized with periodic boundary conditions in a cubic box with coordinate volume $V=L^3$, so that the field can be expanded as
\begin{equation}
    \phi = \sum_{\va*{k}} \left(f_{\va*{k}}A_{\va*{k}} + f^*_{\va*{k}}A^\dagger_{\va*{k}}\right) 
\end{equation}
in terms of some mode functions $f_{\va*{k}}$. After taking expectation values we will let $V$ tend to infinity so that $\va*{k}$ becomes a continuous parameter. From now on, we will also set $\hbar$ to unity and work in comoving time, so that $N=1$ and $\mathcal{H} = \dot{a}/a \equiv H$. 

On imposing the canonical commutation relations
\begin{align}
    \comm{\phi(t,\va*{x})}{\Pi(t,\va*{x}')} &= i\delta(\va*{x} - \va*{x}') \\
    ~
    \comm{A_{\va*{k}}}{A^\dagger_{\va*{k}'}} &= \delta_{\va*{k}\va*{k}'}
\end{align}
we can then interpret $A_{\va*{k}}$ as an annihilation operator and define a vacuum state $\ket{0}$ relative to this mode decomposition by requiring $A_{\va*{k}}\ket{0} = 0$ for all $\va*{k}$. This vacuum definition is dependent on the choice of mode functions, and that choice is in general not unique in generic spacetimes.

\subsection{Early and Late Time Minkowski}
As a first example, we consider the mode functions for a spacetime that asymptotically approaches Minkowski in both the past and the future. 
Given that $a(t)\to a_1$ when $t\to -\infty$ with $a_1$ being a constant, we choose mode functions $f_{\va*{k}}$ that approach the Minkowski vacuum modes \cite{parker_quantum_2009},
\begin{equation}
    f_{\va*{k}} \sim (Va_1^3)^{-1/2}(2\omega_{1k})^{-1/2}e^{i(\va*{k}\vdot\va*{x}-\omega_{1k}t)},
    \label{eq:fmodein}
\end{equation}
in the early time limit, where 
\begin{equation}
    \omega_{1k} = \sqrt{\frac{k^2}{a_1^2} +m^2}, \quad k^2 = \abs{\va*{k}}^2 = \delta_{ij}k^ik^j. 
\end{equation}
The vacuum state $\ket{0}$ defined with respect to these modes is interpreted as the early time vacuum state. By taking the expectation values in (\ref{eq:f1pdef})--(\ref{eq:f3def}) with respect to this vacuum state, and using the mode functions (\ref{eq:fmodein}), the corresponding phase-space functions are 
\begin{align}
f_1^+ &= f_1^- = \frac{1}{2(2\pi)^3},\quad  f_2 = f_3 = 0. 
\end{align}

At late times, $t\to \infty$, we then assume that the spacetime again approaches Minkowski as $a(t)\to a_2$, with $a_2$ a constant. The mode functions $f_{\va*{k}}(t)$ satisfying the early time limit (\ref{eq:fmodein}) will then in general be linear combinations of positive and negative frequency parts \cite{parker_quantum_2009}
\begin{equation}
    f_{\va*{k}} \sim (Va_2^3)^{-1/2}(2\omega_{2k})^{-1/2}e^{i\va*{k}\vdot\va*{x}}\left(\alpha_ke^{-i\omega_{2k}t} + \beta_ke^{i\omega_{2k}t}\right),
    \label{eq:fmodeout}
\end{equation}
with
\begin{equation}
    \omega_{2k} = \sqrt{\frac{k^2}{a_2^2} +m^2}, 
\end{equation}
so that
\begin{equation}
    a_{\va*{k}} = \alpha_k A_{\va*{k}} + \beta^*_k A^\dagger_{-\va*{k}},
\end{equation}
where $a_{\va*{k}}$ is the late time annihilation operator. From this annihilation operator, we find that  the number of outgoing particles in the early time vacuum state is 
\begin{equation}
   \ev{a^\dagger_{\va*{k}}a_{\va*{k}}}{0} = \abs{\beta_k}^2.
\end{equation}
This particle number can be related to the phase-space functions by using the early time vacuum state and (\ref{eq:fmodeout}) in (\ref{eq:f1pdef})--(\ref{eq:f3def}), which leads to 
\begin{align}
f_1^+ &= \frac{1}{2(2\pi)^3}\left(1+2\abs{\beta_k}^2\right), \label{eq:f1p_asymp_flat} \\
f_1^- &= \frac{1}{2(2\pi)^3}, \\
f_2^{\phantom{+}} &= \frac{1}{2(2\pi)^3}\left(\alpha_k\beta^*_k e^{-2i\omega_{2k}t} + \alpha^*_k\beta_k e^{2i\omega_{2k}t} \right), \\
f_3^{\phantom{+}} &= -\frac{i}{2(2\pi)^3}\left(\alpha_k\beta^*_k e^{-2i\omega_{2k}t} - \alpha^*_k\beta_k e^{2i\omega_{2k}t} \right),
\end{align}
for $k^i = \delta^{ij}p_j$. 
Hence we can relate the outgoing particle number to the late time value of $f_1^+$ through 
\begin{equation}
    \abs{\beta_k}^2 = (2\pi)^3f_1^+ - \frac{1}{2}. \label{eq:nk}
\end{equation}
This implies a natural definition of a particle number $n_k$ in terms of $f_1^+$,
\begin{equation}
    n_k \equiv (2\pi)^3f_1^+ - \frac{1}{2}, \label{eq:nk}
\end{equation}
which we extend to intermediate times. This definition can be shown to coincide with the definition in \cite{garbrecht_particle_2004}, where $n_k$ in a flat FLRW spacetime was found by determining the Bogoliubov transformation that gave the maximum number of particles.

\subsection{de Sitter Inflation} 
As a second example, now consider (half of) de Sitter spacetime with flat spatial slicings and constant Hubble parameter $H$. Defining the vacuum state $\ket{0}$ as the Bunch-Davies vacuum, the corresponding  mode functions are \footnote{This expression is based on Ref.~\cite{parker_quantum_2009}, but we have added a factor $e^{-\pi\Im(\nu)/2}$ to get consistent normalization for imaginary $\nu$. Up to an unimportant constant overall phase factor, these mode functions have the same form as the in-vacuum modes used in Refs.~\cite{anderson_instability_2014,anderson_decay_2018} for flat FLRW coordinates.}
\begin{equation}
f_{\va*{k}}=\frac{1}{2}\sqrt{\frac{\pi}{HV}}e^{-3Ht/2}e^{-\pi\Im(\nu)/2}H_{\nu}^{(1)}\left(\frac{k}{H}e^{-Ht}\right)e^{i\va*{k}\vdot\va*{x}},
\end{equation}
where $H_{\nu}^{(1)}$ is a Hankel function and 
\begin{equation}
    \nu = \sqrt{\frac{9}{4} - \frac{m^2}{H^2}}.
\end{equation}
 The distribution functions corresponding to this vacuum state are in turn given by
\begin{widetext}
\begin{align}
    f_1^+ &= \frac{e^{-\pi\Im(\nu)}}{2(2\pi)^3}\frac{\pi}{4H}\omega_p \left[ \abs{H_{\nu}^{(1)}}^2 + \frac{p^2}{a^2\omega_p^2}\abs{\left(H_{\nu}^{(1)}\right)' +\frac{3}{2\zeta}H_{\nu}^{(1)}}^2\right],\label{eq:f1pdesitter} \\
    ~
    f_1^- &=-\frac{e^{-\pi\Im(\nu)}}{(2\pi)^3}\frac{\pi\zeta}{4}\Im{H_{\nu}^{(1)}\left(\left(H_{\nu}^{(1)}\right)' +\frac{3}{2\zeta}H_{\nu}^{(1)} \right)^*} = \frac{1}{2(2\pi)^3}, \\
    ~
    f_2^{\phantom{+}} &= \frac{e^{-\pi\Im(\nu)}}{2(2\pi)^3}\frac{\pi}{4H}\omega_p \left[ \abs{H_{\nu}^{(1)}}^2 - \frac{p^2}{a^2\omega_p^2}\abs{\left(H_{\nu}^{(1)}\right)' +\frac{3}{2\zeta}H_{\nu}^{(1)}}^2\right], \\
    ~
    f_3^{\phantom{+}} &= -\frac{e^{-\pi\Im(\nu)}}{(2\pi)^3}\frac{\pi\zeta}{4}\Re{H_{\nu}^{(1)}\left(\left(H_{\nu}^{(1)}\right)' +\frac{3}{2\zeta}H_{\nu}^{(1)} \right)^*}, \label{eq:f3desitter} 
\end{align}
\end{widetext}
where $p^2 \equiv \delta^{ij}p_ip_j$ and the Hankel functions should be evaluated at $\zeta \equiv p/(aH)$. A prime denotes differentiation with respect to $\zeta$. 
Using the properties of the Hankel functions, it can be shown that (\ref{eq:f1pdesitter})--(\ref{eq:f3desitter}) is indeed a solution to (\ref{eq:f1peqFLRW})--(\ref{eq:f3eqFLRW}) for the de Sitter spacetime \footnote{The fact that $f_1^-$ is constant and equal to $1/(2(2\pi)^3)$ in all of our applications is related to the normalization of the mode functions and the conservation of the Wronskian (see e.g.~Ref.~\cite{parker_quantum_2009}).  }.

To compare these results with other references, we consider some specific values of the parameters $m$ and $\nu$. First, for $m=0$, $\nu = 3/2$, the distribution functions can be simplified to
\begin{align}
f_1^+ &=\frac{1}{2(2\pi)^3}\left(1+\frac{H^2a^2}{2p^2}\right),\\
~
f_1^- &= \frac{1}{2(2\pi)^3}, \\
f_2^{\phantom{+}} &= \frac{1}{2(2\pi)^3}\frac{H^2a^2}{2p^2}, \\
f_3^{\phantom{+}} &= -\frac{1}{2(2\pi)^3}\frac{Ha}{p}, 
\end{align}
giving
\begin{equation}
    n_k = \frac{H^2a^2}{4p^2},
    \label{eq:nk_massless}
\end{equation}
which coincides with the result found in \cite{garbrecht_particle_2004}. Note, however, that this particle number will give an infinite result upon integrating over the momenta.

As a second example, we compare with the results in \cite{anderson_instability_2014,anderson_decay_2018} for imaginary orders $\nu$. Defining $\nu \equiv i\bar{\gamma}$, with $\bar{\gamma}$ now being real, and taking the limit $\zeta \to 0$, corresponding to $t \to \infty$, we get 

\begin{align}
f_1^+ = {}&
\frac{1}{2(2\pi)^3}\bigg(\frac{m}{H\bar{\gamma}}\coth(\pi\bar{\gamma}) \notag
 \\
 &+\frac{3}{2\bar{\gamma}}\csch(\pi\bar{\gamma})\cos\left(2\bar{\gamma} Ht+\psi\right) \bigg), 
\label{eq:f1p_dS_late}\\
~
f_1^- ={}& \frac{1}{2(2\pi)^3}, \\
f_2^{\phantom{+}} ={}& -\frac{1}{2(2\pi)^3}\csch(\pi \bar{\gamma})\sin(2\bar{\gamma} Ht + \psi), \\
f_3^{\phantom{+}} = {}&-\frac{1}{2(2\pi)^3}\bigg(\frac{3}{2\bar{\gamma}}\coth(\pi\bar{\gamma}) \notag\\
&+\frac{m}{H\bar{\gamma}}\csch(\pi\bar{\gamma})\cos\left(2\bar{\gamma} Ht+\psi\right) \bigg),
\label{eq:f3_dS_late}
\end{align}
where the phase $\psi$ is given by
\begin{equation}
    \psi = 2\arg(\Gamma(i\bar{\gamma}))-\arctan(\frac{2\bar{\gamma}}{3})-2\bar{\gamma}\ln(\frac{p}{2H}).
\end{equation}
Using the same particle definition as previously, we find
\begin{align}
     n_k = {}& \frac{m}{2H\bar{\gamma}}\left(1+\frac{H\bar{\gamma}}{m}\right)\frac{e^{-2\pi\bar{\gamma}}}{1-e^{-2\pi\bar{\gamma}}} \notag \\
     &+\frac{m}{2H\bar{\gamma}}\left(1-\frac{H\bar{\gamma}}{m}\right)\frac{1}{1-e^{-2\pi\bar{\gamma}}} \notag \\
     &+\frac{3}{2\bar{\gamma}}\frac{e^{-\pi\bar{\gamma}}}{1-e^{-2\pi\bar{\gamma}}}\cos(2\bar{\gamma} Ht+\psi). \label{eq:nk_massive_asymptotic}
\end{align}
The first row of this expression gives a similar contribution as in \cite{anderson_decay_2018}, where the particle number for the Bunch-Davies in-vacuum relative to the asymptotic adiabatic out vacuum was found to be $e^{-2\pi\bar{\gamma}}/(1-e^{-2\pi\bar{\gamma}})$ \footnote{ Note that the precise value of $\bar{\gamma}$ for a specified mass $m$ is slightly different  in Ref.~\cite{anderson_decay_2018} due to the conformal coupling used there.}. This result agrees with the first line in (\ref{eq:nk_massive_asymptotic}) to leading order in the $m\gg H$ limit, where both reduce to $e^{-2\pi m/H}$. However, due to the prefactors and the following rows, the particle number defined here differs from the adiabatic result in \cite{anderson_decay_2018}. 

\section{Relation to the Adiabatic Particle Number}
To more clearly see the difference between $n_k$ and the commonly used adiabatic definitions, we now write the adiabatic particle numbers in terms of the kinetic phase-space functions. 
For this purpose, we first  define the adiabatic mode functions
\begin{equation}
    \Tilde{f}_{\va*{k}} = \frac{e^{i\va*{k}\vdot \va*{x}}}{\sqrt{2Va^3W_k(t)}}e^{-i\Theta_k(t)},
    \label{eq:adiabatic_modes}
\end{equation}
where
\begin{equation}
    \Theta_k = \int^t \dd{t'} W_k(t'),
\end{equation}
is the adiabatic phase \cite{habib_energy-momentum_1999}. If we would assume that the exact scalar field has the same form as these mode functions, the equations of motion for the field could then be written as a differential equation for $W_k$. Expanding this equation in powers of time derivatives acting on the scale factor, we would then obtain an expression for $W_k$ order by order. This expansion procedure is usually referred to as the adiabatic expansion of the field. However, if we do not require that the scalar field has this precise form in terms of $W_k$, it is not necessary to require that the mode functions (\ref{eq:adiabatic_modes}) satisfy the equations of motion. Instead, these modes will here rather be thought of as serving as a basis that we can compare the exact mode functions to, without assuming that the basis modes satisfy the dynamical equations. The exact mode functions can then be written in terms of the basis modes through a time dependent Bogoliubov transformation
\begin{equation}
    f_{\va*{k}} = \alpha_k(t)\Tilde{f}_{\va*{k}} + \beta_k(t)\Tilde{f}_{-\va*{k}}^*.
\end{equation}
We then define a function $V_k(t)$ by requiring that \cite{habib_energy-momentum_1999} 
\begin{align}
    \dot{f}_{\va*{k}} = {}&\left(-iW_k + \frac{V_k}{2}-\frac{3H}{2}\right)\alpha_k \Tilde{f}_{\va*{k}}\notag \\
    ~
    &+ \left(iW_k + \frac{V_k}{2}-\frac{3H}{2}\right)\beta_k \Tilde{f}_{-\va*{k}}^*.
\end{align}
Both $W_k(t)$ and $V_k(t)$  are here assumed to be real. Since the adiabatic basis functions are not required to satisfy the equations of motion, there is some freedom in choosing the functions $W_k(t)$ and $V_k(t)$ that define the basis. A physically motivated choice is, however, to choose $W_k$ and $V_k$ to match  divergences in the adiabatic expansion of the exact mode functions when calculating the energy-momentum tensor \cite{anderson_instability_2014}.    

Writing the phase-space functions in terms of $W_k$, $V_k$, $\alpha_k$, $\beta_k$, and $\Theta_k$, we see that
\begin{align}
    f_1^+ ={}& \frac{1}{2(2\pi)^3}\frac{\omega_p}{2W_k}\Bigg[ \left(1+\frac{A_w^2}{\omega_p^2}\right)\left(1+2\abs{\beta_k}^2\right) \notag \\
    &+ 2\Re{\alpha_k^*\beta_ke^{2i\Theta_k}\left(1+\frac{A_w^2e^{2i\delta_w}}{\omega_p^2}\right)}\Bigg],
  \label{eq:f1padiabatic}\\
    ~
    f_1^- ={}& \frac{1}{2(2\pi)^3}, \\
    ~
    f_2^{\phantom{+}} ={}& 
    \frac{1}{2(2\pi)^3}\frac{\omega_p}{2W_k}\Bigg[ \left(1-\frac{A_w^2}{\omega_p^2}\right)\left(1+2\abs{\beta_k}^2\right) \notag \\
    &+ 2\Re{\alpha_k^*\beta_ke^{2i\Theta_k}\left(1-\frac{A_w^2e^{2i\delta_w}}{\omega_p^2}\right)}\Bigg], \\
    ~
    f_3^{\phantom{+}} = {}&
    \frac{1}{2(2\pi)^3}\frac{1}{W_k}\Bigg[\left(\frac{V_k}{2}-\frac{3H}{2}\right)\left(1+2\abs{\beta_k}^2\right) \notag \\
    &+2A_w\Re{\alpha_k^*\beta_ke^{2i\Theta_k}e^{i\delta_w}}\Bigg], 
    \label{eq:f3adiabatic}
\end{align}
where
\begin{align}
    A_w &= \abs{iW_k + \frac{V_k}{2}-\frac{3H}{2}}, \\
    \delta_w  &= \arg\left(iW_k + \frac{V_k}{2}-\frac{3H}{2}\right), 
\end{align}
and $k^i = \delta^{ij}p_j$. Note that we can write
\begin{widetext}
\begin{align}
\Re{\alpha_k^*\beta_ke^{2i\Theta_k}\left(1\pm\frac{A_w^2e^{2i\delta_w}}{\omega_p^2}\right)} &= \mathcal{R}_k(t)\left(1 \mp \frac{4W_k^2 - (V_k-3H)^2}{4\omega_p^2}\right) 
\pm \mathcal{I}_k(t)\frac{W_k(V_k-3H)}{\omega_p^2}, \\
A_w\Re{\alpha_k^*\beta_ke^{2i\Theta_k}e^{i\delta_w}} &=\mathcal{R}_k(t)\left(\frac{V_k}{2}-\frac{3H}{2}\right) + W_k\mathcal{I}_k(t), 
\end{align}
where
\begin{align}
       \mathcal{R}_k(t) &\equiv \Re{\alpha_k\beta_k^*e^{-2i\Theta_k}} = \Re{\alpha_k^*\beta_ke^{2i\Theta_k}}, \\
    ~
    \mathcal{I}_k(t) &\equiv \Im{\alpha_k\beta_k^*e^{-2i\Theta_k}} = -\Im{\alpha_k^*\beta_ke^{2i\Theta_k}}, 
\end{align}
are oscillatory quantum interference functions  \cite{habib_energy-momentum_1999, anderson_instability_2014}.

On combining some of the phase-space functions, we can extract a particle number by noting that
\begin{align}
\mathcal{R}_k = {}& (2\pi)^3\frac{W_k}{\omega_p}\left(f_1^+ +f_2\right)-\abs{\beta_k}^2 -\frac{1}{2},\\
\mathcal{I}_k = {}&(2\pi)^3\left(f_3-\left(\frac{V_k}{2}-\frac{3H}{2}\right)\frac{f_1^+ + f_2}{\omega_p}\right),
\end{align}
so that 
\begin{align}
    \mathcal{N}_k \equiv & \abs{\beta_k}^2   = (2\pi)^3\frac{\omega_p}{2W_k} \left(2f_1^+ -\left(f_1^+ +f_2\right)\left(1 - \frac{A_w^2}{\omega_p^2}\right) - f_3\frac{(V_k-3H)}{\omega_p}\right) - \frac{1}{2}, \label{eq:adiabatic_N_in_terms_of_f}
\end{align}
\end{widetext}
giving a definition of the adiabatic particle number $\mathcal{N}_k$ in terms of the phase-space functions, $W_k$, and $V_k$. Depending on how $W_k$ and $V_k$ are chosen, and up to which order they match the adiabatic expansion of the exact solutions, we get different particle numbers. In the following we will use the collective term \textit{adiabatic} particle numbers for all $\mathcal{N}_k$ obtained on choosing $W_k$ and $V_k$ to match the adiabatic expansion up to some order, but it should be noted that this term is in some references reserved for the particle number to lowest adiabatic order. To zeroth order, with $W_k = \omega_p$, $V_k = 0$, and neglecting the term explicitly involving $H$, we see that $\mathcal{N}_k$ reduces to $n_k$. Hence $n_k$ can be seen as a zeroth order adiabatic particle number.

To see the effect of the relation between the phase-space functions and the adiabatic particle number, we can consider the late time de Sitter case. For this purpose we use adiabatic functions that are correct up to first adiabatic order 
\begin{align}
    W_k^{(0)} &= \sqrt{\omega_p^2-\frac{9H^2}{4}}, \label{eq:WkMottola}\\
    ~
    V_k^{(1)} &= -\frac{\dot{\omega}_p}{\omega_p} = H\left(1-\frac{m^2}{\omega_p^2}\right), \label{eq:VkMottola}
\end{align}
which are similar to functions used in \cite{anderson_instability_2014, anderson_decay_2018}.
Inserting these into (\ref{eq:adiabatic_N_in_terms_of_f}) in the $\zeta \to 0$ limit and using (\ref{eq:f1p_dS_late})--(\ref{eq:f3_dS_late}) then gives
\begin{align}
    \mathcal{N}_k^{(1)}  &= (2\pi)^3\frac{m}{\bar{\gamma} H} \left(f_1^+  + \frac{3H}{2m}f_3 \right)- \frac{1}{2}  \notag \\
    &= \frac{1}{2}\coth(\pi\bar{\gamma})- \frac{1}{2},
\end{align}
where the superscript on $\mathcal{N}_k$ denotes that this adiabatic particle number was obtained from (\ref{eq:adiabatic_N_in_terms_of_f}) with the choice (\ref{eq:WkMottola})--(\ref{eq:VkMottola}) of $W_k$ and $V_k$.  Thus we see that, for this choice of $W_k$ and $V_k$,  the oscillations in the phase-space functions cancel, and we are left with a result of the same form as in \cite{anderson_decay_2018}.

\section{Numerical results}
Having described the connection between the adiabatic particle numbers $\mathcal{N}_k$ and the particle number $n_k$ given in terms of $f_1^+$, we proceed with a numerical investigation of how their dynamics differ in practice for some prescribed scale factor profiles and initial conditions.

\subsection{Prescribed profiles}
The scale factor profiles we will consider are those for de Sitter inflation, de Sitter inflation with a cut-off at time $t_c$, a finite $H$ pulse, and a dust cosmology. These are represented by 
\begin{align}
    a(t) &= e^{Ht}, \\
    a(t) &= a_N\exp \Bigg\{ \frac{At}{2} + \frac{AW}{2\pi}\Bigg(\ln((t - t_c)^2 + W^2) \notag\\
    &- \frac{2(t - t_c)}{W}\arctan(\frac{t - t_c}{W})\Bigg) \Bigg\}, \label{eq:a_cutoff} \\ 
    a(t) &= \exp\left\{2WA\arctan\left(\exp\left(\frac{t-t_p}{W}\right)\right)\right\}, \label{eq:a_pulse}\\
    a(t) &= t^{2/3},
\end{align}
respectively, where (\ref{eq:a_cutoff}) and (\ref{eq:a_pulse}) correspond to
\begin{align}
    H(t) &= A\left(\frac{1}{2}-\frac{1}{\pi}\arctan(\frac{t-t_c}{W})\right), \\
    H(t) &= A\sech(\frac{t-t_p}{W}),
\end{align}
and where $A$, $W$, $t_c$, $t_p$, and $a_N$ are constants.
$H$ is constant in the de Sitter case, and the factor $a_N$ in the cut-off case normalizes $a$ to unity at $t=0$.   

\subsection{Initial conditions}
The initial conditions for the phase-space functions are set to correspond to an initial adiabatic vacuum. This initial data is obtained by setting $\abs{\beta_k}=0$ in Eqs.~(\ref{eq:f1padiabatic})--(\ref{eq:f3adiabatic}), which gives
\begin{align}
    f_1^+ &= \frac{1}{2(2\pi)^3}\frac{\omega_p}{2W_k}\left(1+\frac{A_w^2}{\omega_p^2}\right), \\
    ~
    f_1^- &= \frac{1}{2(2\pi)^3}, \\
    ~
    f_2^{\phantom{+}} &=  \frac{1}{2(2\pi)^3}\frac{\omega_p}{2W_k}\left(1-\frac{A_w^2}{\omega_p^2}\right),\\
    ~
    f_3^{\phantom{+}} &= \frac{1}{2(2\pi)^3}\frac{1}{W_k}\left(\frac{V_k}{2}-\frac{3H}{2}\right),
\end{align}
where all quantities should be evaluated at the initial time. For the adiabatic functions appearing in the initial data, we use a set that is correct up to fourth adiabatic order
\begin{align}
    &W_k = \omega_p+\frac{R\Bar{\xi}}{2\omega_p} -\frac{m^2\left(3H^2+\dot{H}\right)}{4{\omega_p}^3}+\frac{5H^2m^4}{8{\omega_p}^5} \notag\\
    ~
    &-\frac{R^2{\Bar{\xi}}^2}{8{\omega_p}^3}-\frac{\Bar{\xi}\left(6RH^2+5\dot{R}H+\ddot{R}+2\dot{H}R\right)}{8{\omega_p}^3}\notag \\
    ~
    &+\frac{m^2\left(60H^4+86H^2\dot{H}+15\ddot{H}H+10{\dot{H}}^2+\dddot{H}\right)}{16{\omega_p}^5} \notag\\
    ~
    &+\frac{m^2\Bar{\xi}\left(19RH^2+5\dot{R}H+3\dot{H}R\right)}{8{\omega_p}^5} -\frac{25H^2Rm^4\Bar{\xi}}{16{\omega_p}^7}\notag\\
    ~
    &-\frac{m^4\left(507H^4+394H^2\dot{H}+28\ddot{H}H+19{\dot{H}}^2\right)}{32{\omega_p}^7}\notag\\
    ~
    &+\frac{221H^2m^6\left(3H^2+\dot{H}\right)}{32{\omega_p}^9}-\frac{1105H^4m^8}{128{\omega_p}^{11}} \notag\\
    ~
    &+\sqrt{\omega_p^2-\frac{9H^2}{4}} - \omega_p\left(1-\frac{9H^2}{8\omega_p^2}-\frac{81H^4}{128\omega_p^4}\right),
    \label{eq:Wk4thorder}
\end{align}

\begin{align}
&V_k =H\left(1-\frac{m^2}{{\omega_p}^2}\right) -\frac{\Bar{\xi}\left(\dot{R}+2HR\right)}{2{\omega_p}^2} \notag \\
    ~
    &+ \frac{m^2\left(12H^3+10\dot{H}H+\ddot{H}\right)}{4{\omega_p}^4}+\frac{HRm^2\Bar{\xi}}{{\omega_p}^4}
    \notag \\
    ~
    &+\frac{15H^3m^6}{4{\omega_p}^8} -\frac{9Hm^4\left(3H^2+\dot{H}\right)}{4{\omega_p}^6},
    \label{eq:Vk4thorder}
\end{align}
where
\begin{equation}
    R = 6\left(\left(\frac{\dot{a}}{a}\right)^2 + \frac{\ddot{a}}{a}\right) = 6\left(\dot{H}+2H^2\right),
\end{equation}
and $\Bar{\xi}=-1/6$. This choice is based on a fourth order result presented in Ref.~\cite{habib_energy-momentum_1999}, but we have added the square root and the last three terms in $W_k$ \footnote{We have also corrected some apparent misprints in Ref.~\cite{habib_energy-momentum_1999}. For instance, our $V_k$ should here correspond to the expression for $-\dot{W}_k^{(2)}/W_k^{(2)}$ in Ref.~\cite{habib_energy-momentum_1999}, but some of the numerical coefficients are different. We have also corrected for a missing $m^2$ in $W_k$ }. By construction, these terms only modify the expression with terms of adiabatic order six and higher, maintaining a $W_k$ that is correct up to fourth order. The reason for the addition is to avoid introducing unwanted oscillations in the late-time de Sitter case when calculating the adiabatic particle number through (\ref{eq:adiabatic_N_in_terms_of_f}).

\subsection{Numerical particle numbers}
\begin{figure*}[pt]
    \centering
    \begin{subfigure}[t]{0.5\textwidth}
        \centering
        \includegraphics[scale = 0.5]{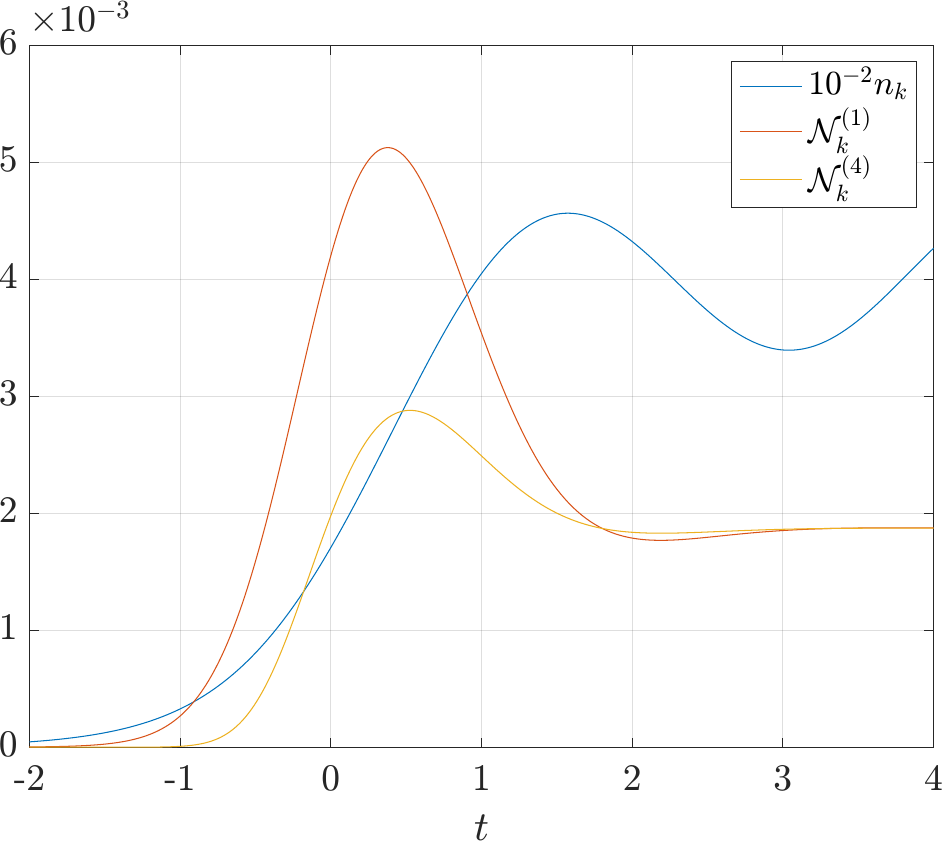}
        \caption{}
    \end{subfigure}%
    ~ 
    \begin{subfigure}[t]{0.5\textwidth}
        \centering
        \includegraphics[scale = 0.5]{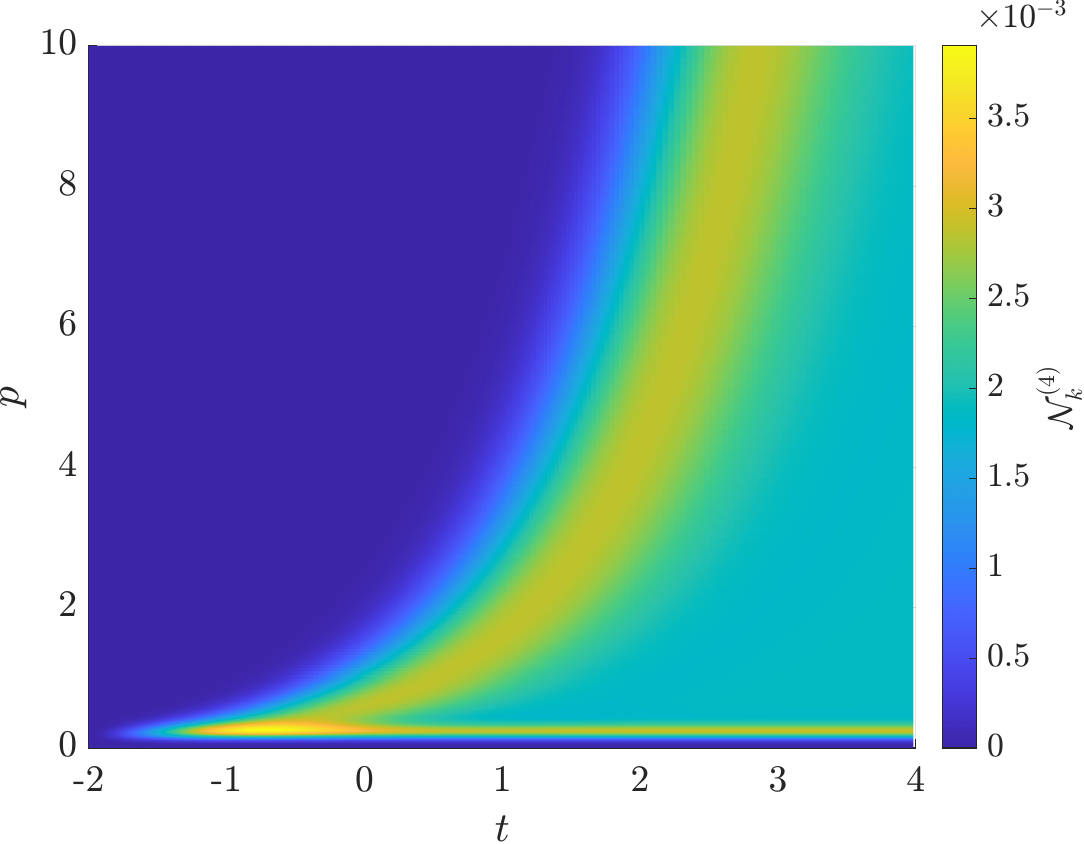}
        \caption{}
    \end{subfigure}

    \begin{subfigure}[t]{0.5\textwidth}
        \centering
        \includegraphics[scale = 0.5]{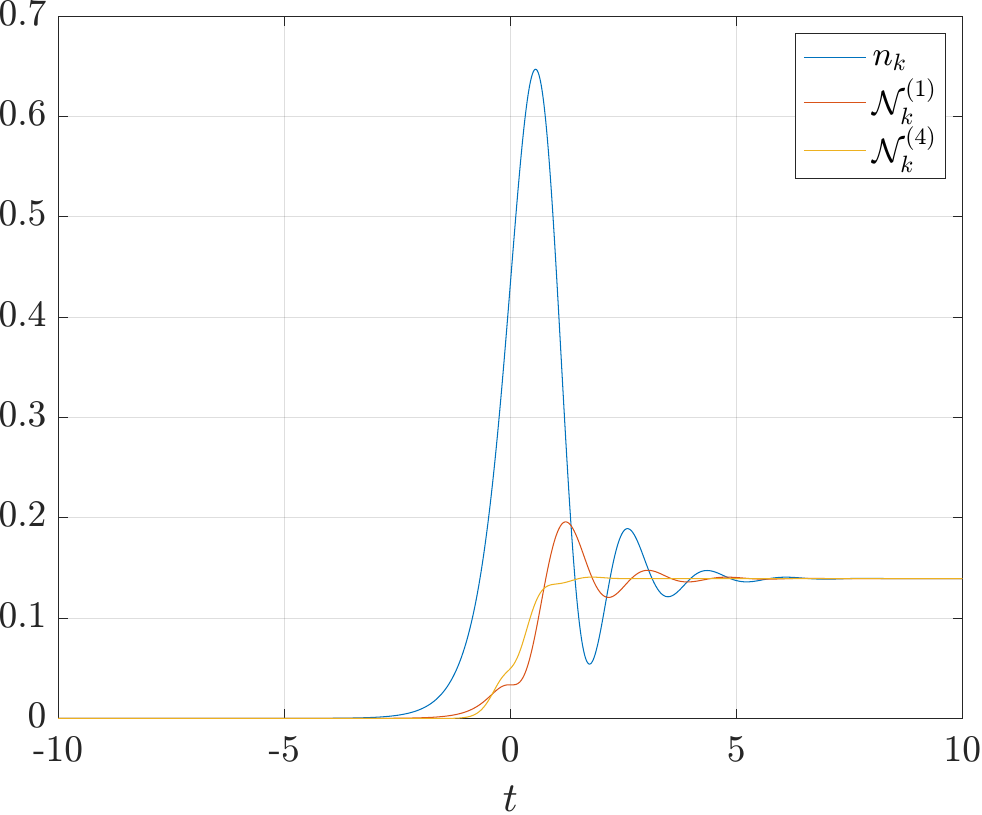}
        \caption{}
    \end{subfigure}%
    ~ 
    \begin{subfigure}[t]{0.5\textwidth}
        \centering
        \includegraphics[scale = 0.5]{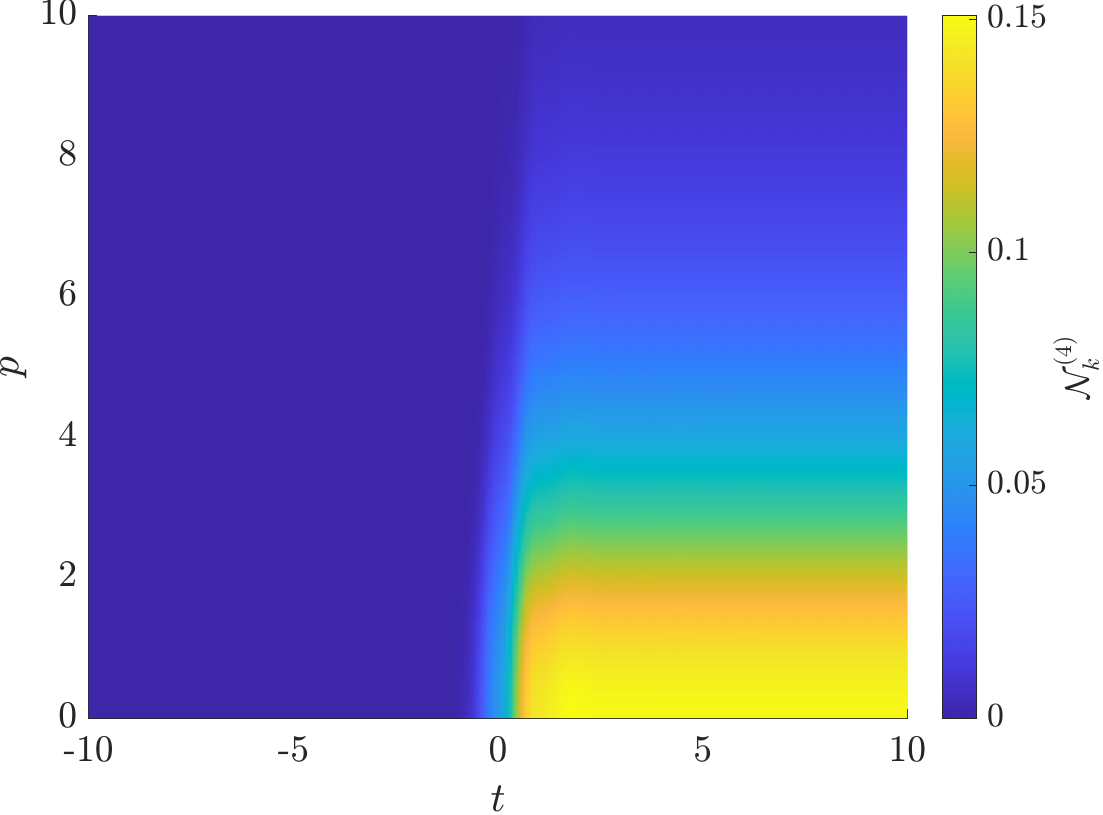}
        \caption{}
    \end{subfigure}

    \begin{subfigure}[t]{0.5\textwidth}
        \centering
        \includegraphics[scale = 0.5]{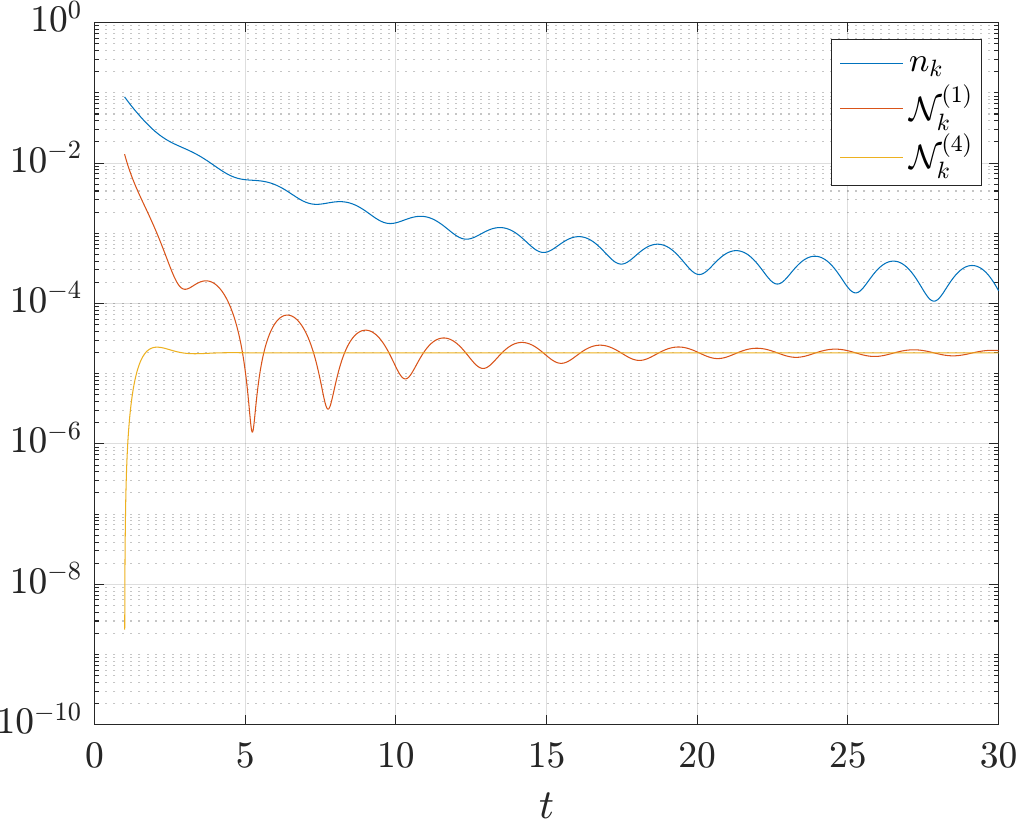}
        \caption{}
    \end{subfigure}%
    ~ 
    \begin{subfigure}[t]{0.5\textwidth}
        \centering
        \includegraphics[scale = 0.5]{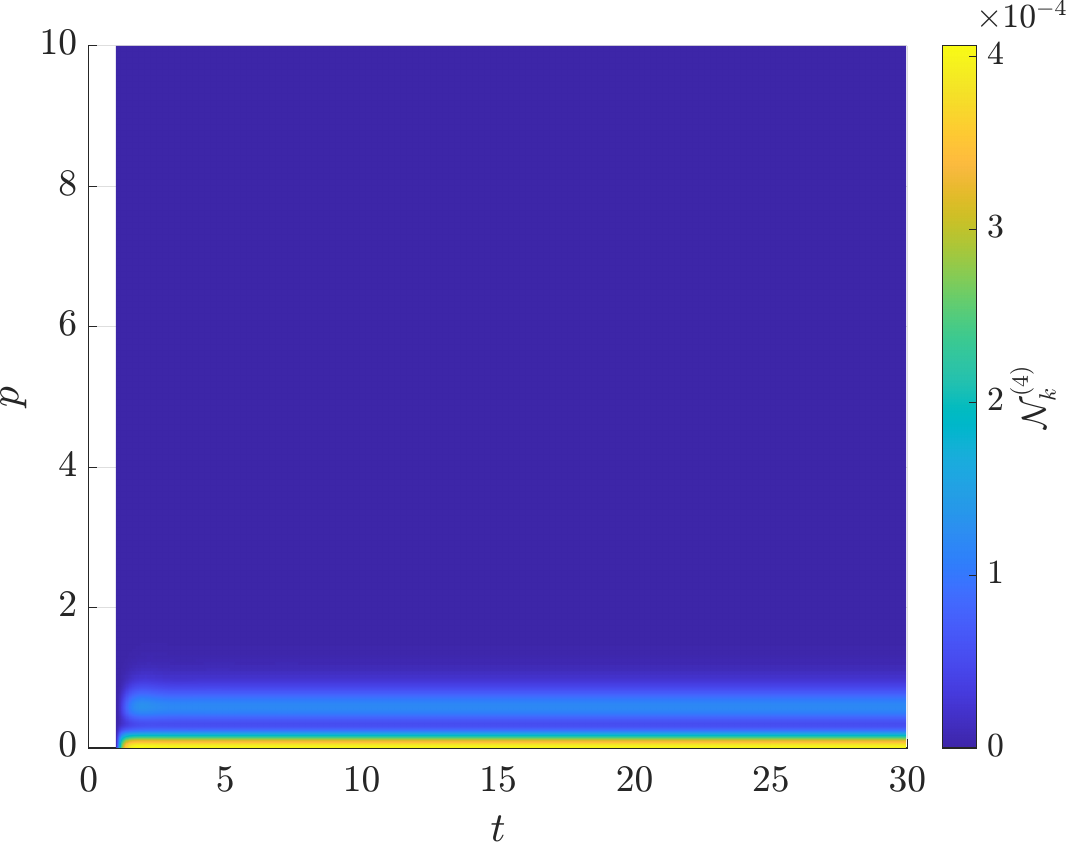}
        \caption{}
    \end{subfigure}
    \caption{Particle numbers for different scale factor profiles starting from a fourth order adiabatic vacuum. The profiles are \textbf{a--b)} de Sitter inflation with $H=1$, \textbf{c--d)} a $\sech$ pulse in $H$ with $t_p = 0$ and $W=A=1$, \textbf{e--f)} a dust cosmology. In each of these the mass is set to $m = \sqrt{13}H_{\text{max}}/2$ where $H_{\text{max}}$ is the maximum value during the evolution. The panels on the left hand side are for $p=1$. Note that $n_k$ is multiplied by $10^{-2}$ in a)}
    \label{fig:3x2ParticleNumberPlot}
\end{figure*}

In Fig.~\ref{fig:3x2ParticleNumberPlot} we show particle numbers for de Sitter inflation, the $\sech$ pulse, and the dust cosmology when starting with an initial fourth order adiabatic vacuum. The adiabatic particle number obtained by using (\ref{eq:WkMottola})--(\ref{eq:VkMottola}) in (\ref{eq:adiabatic_N_in_terms_of_f}) is denoted by $\mathcal{N}_k^{(1)}$, while $\mathcal{N}_k^{(4)}$ is the number found when using (\ref{eq:Wk4thorder})--(\ref{eq:Vk4thorder}).

In the de Sitter case we see that each mode has a distinct creation event, with higher momenta being created later. After being created, the adiabatic paritcle numbers stabilize without any oscillations. This was seen previously in \cite{anderson_decay_2018,anderson_instability_2014}. However, the particle number $n_k$ continues to oscillate, and goes up to much larger values than the adiabatic counterparts. 

Looking at the pulse and dust cases, we see that the particle creation is concentrated to smaller momenta, with the creation being centered in time around the maximum of the pulse and at the beginning of the dust evolution respectively. When the expansion subsides and tends to zero, the difference between $n_k$ and the adiabatic particle numbers becomes smaller, and the oscillations in $n_k$ decrease in amplitude. 
\begin{figure*}[t!]
    \centering
    \begin{subfigure}[t]{0.5\textwidth}
        \centering
        \includegraphics[scale = 0.5]{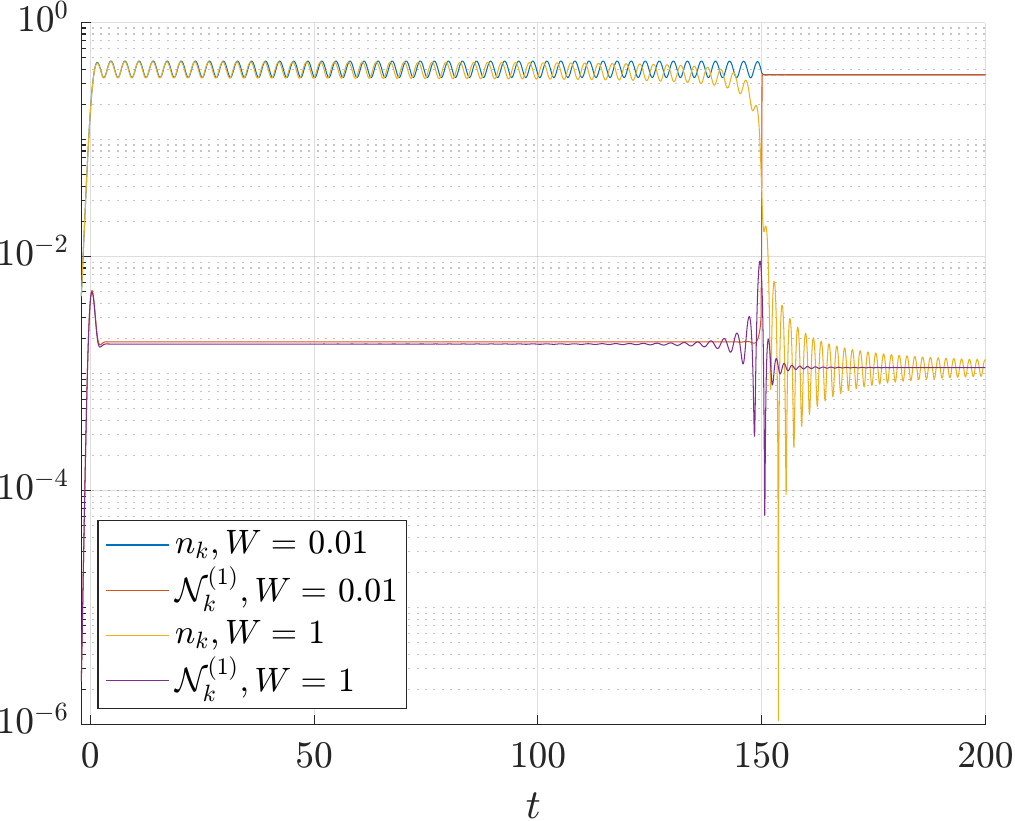}
        \caption{}
    \end{subfigure}%
    ~ 
    \begin{subfigure}[t]{0.5\textwidth}
        \centering
        \includegraphics[scale = 0.5]{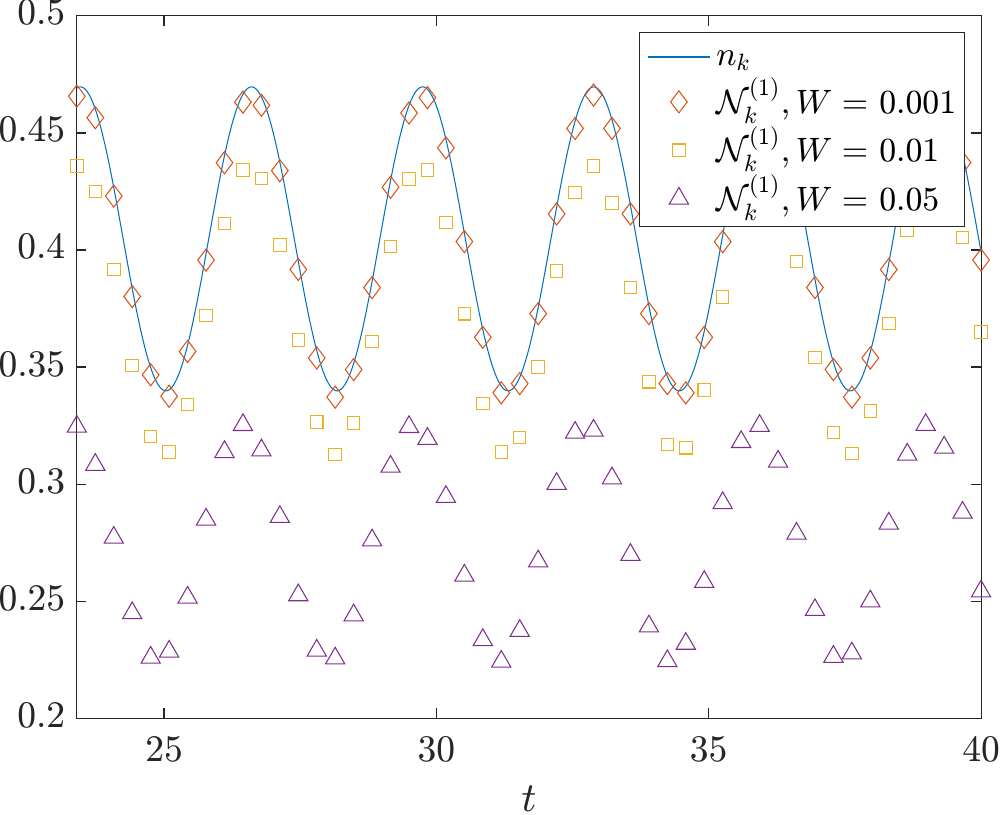}
        \caption{}
    \end{subfigure}
    \caption{Particle numbers for $p = 1$ starting from a fourth order adiabatic vacuum for de Sitter inflation with a cut-off. \textbf{a)} shows the behaviour for $t_c = 150$ but with different sizes of the transition region $W$. \textbf{b)} shows $n_k$ without a cut-off while the markers show the $\mathcal{N}_k^{(1)}$ obtained asymptotically after the expansion is switched off at time $t$.    }
    \label{fig:SwitchOff}
\end{figure*}

To more clearly see where the difference in the particle numbers originate from, we can consider the case when the de Sitter inflation is cut off at some time $t_c$. This is shown in Fig.~\ref{fig:SwitchOff} for different values of the size of the transition region $W$ and cut-off time $t_c$. We have here chosen to omit the fourth order adiabatic particle number, as this number goes through very large swings in the transition region. These are due to the derivatives of the Hubble parameter becoming very large for small $W$.

In Fig.~\ref{fig:SwitchOff}, we see that the end result will depend on how fast the cut-off is chosen to be and where it is centered. If the cut-off is very fast, corresponding to a small transition region $W$, $\mathcal{N}_k^{(1)}$ shoots up to match $n_k$. In turn, $n_k$ stays at about the same value as before the cut-off, but without any oscillations after the cut-off. If we instead make the transition region larger, we see that $n_k$ will decrease down towards $\mathcal{N}_k^{(1)}$, but the asymptotic result does not perfectly match the $\mathcal{N}_k^{(1)}$ plateau before the switch, at least for the chosen $W$.  

The precise value at which the numbers stabilize is also found to depend on where the cut-off is centered. This can be seen in Fig.~\ref{fig:SwitchOff}b, where the placement of $t_c$ in relation to the oscillations in $n_k$ lead to different asymptotic particle numbers.  The asymptotic numbers are displayed as markers in the figure, where a marker at time $t$ corresponds to the asymptotic particle number obtained on choosing $t_c = t$ for that specific value of $t$. As seen in the figure, the particle numbers obtained asymptotically after a switch-off at time $t$ follow the same shape as $n_k(t)$ calculated without the switch, and approach $n_k(t)$ as the transition region is made smaller. 

Based on these observations, $n_k$ can be interpreted as the particle number that would be obtained if the expansion rate is very rapidly switched off, whereas $\mathcal{N}_k^{(1)}$ is closer to the value obtained asymptotically for a slow adiabatic switch-off. The interpretation of $n_k$ can also be seen through the differential equation (\ref{eq:f1peqFLRW}), where quickly switching off $H$ would result in $f_1^+$, and hence $n_k$, becoming frozen in at the value it had just before the switch. Since spacetime is Minkowski after the switch, this value must directly correspond to the particle number through (\ref{eq:f1p_asymp_flat}).

With this interpretation, $n_k$ behaves in a similar fashion as the adiabatic particle number studied in \cite{ilderton_physics_2022} in the context of flat spacetime QED. The term adiabatic particle number was there used to mean the particle number relative to instantaneous eigenstates of the Hamiltonian, corresponding to a zeroth order adiabatic particle number in our terminology. Since $n_k$ can also be seen as a zeroth order adiabatic particle number, and furthermore corresponds to a Bogoliubov transformation that diagonalizes the Hamiltonian density \cite{garbrecht_particle_2004}, the similarities between $n_k$ and the particle number in \cite{ilderton_physics_2022} were expected. A similar interpretation was also made in a kinetic QED context using the Wigner formalism in \cite{diez_identifying_2023}.

\section{\label{sec:Regularization}Regularization}
In flat spacetime QED, the total particle density after the rapid switch-off gives a finite result when integrated over the momenta, which strengthens the interpretation of the particle number relative to instantaneous Hamiltonian eigenstates as something potentially accessible \cite{ilderton_physics_2022}. However, as hinted at in \cite{ilderton_physics_2022}, in the gravitational scenario we are faced with the problem that $n_k$ in general needs to be regularized to give a finite result when integrated over the momentum space. In cosmology, this is often done using an adiabatic subtraction scheme. To regularize the energy density,
\begin{equation}
    \rho = \frac{4\pi}{a^3}\int \dd{p}p^2\omega_p f_1^+,
\end{equation}
we generally have to subtract terms up to fourth order in the adiabatic expansion. To deduce the subtractions needed for the particle number $n_k$, which is given in terms of $f_1^+$, it suffices to look at this subtraction up to second order \cite{moreno-pulido_renormalizing_2022},   
\begin{align}
    \rho -\tensor[^{(0-2)}]{\rho}{} =&{} \frac{1}{a^3}\int \frac{\dd{p}}{(2\pi)^3}4\pi p^2\omega_p \Bigg[(2\pi)^3f_1^+ - \frac{1}{2} \notag \\ &-\frac{H^2}{4\omega_p^2}-\frac{m^2H^2}{4\omega_p^4} -\frac{m^4H^2}{16\omega_p^6}\Bigg],
    \label{eq:rhoren2}
\end{align}
since the fourth order divergences become finite when dividing with an extra $\omega_p$. The first two terms inside the bracket combine to form the bare $n_k$ defined earlier. As for the third term, looking at the massless case we see that this term precisely subtracts the divergent value of $n_k$ that we found for the de Sitter spacetime earlier. Finally, the last two terms would correspond to finite second order subtractions on the particle number level.

The general necessity of regularizing $n_k$ together with its interpretation as the particle number obtained after a rapid switch-off implies that, if, for whatever reason, the expansion of the universe could be suddenly switched off, that would in general lead to the creation of an infinite particle density. Hence,  even in theory, the potential accessibility of this particle number is questionable.  However, the interpretation of $n_k$ as the particle number that would be obtained after the switch is still valid.

\section{Conclusions}

We have shown how the quantum kinetic formalism from Ref.~\cite{friedrich_kinetic_2018} can be used to study particle production in cosmology. Thinking in terms of a hypothetical switch-off in the cosmological expansion rate, we have given a clear interpretation of a key particle definition, $n_k$, as the number of particles that would be obtained after the switch. However, when working in an expanding universe, the total number of particles that we obtain this way turns out to be infinite and does therefore not
correspond to physically accessible particles. Nonetheless, the interpretation of this particle number is still valid and gives a clear and unambiguous meaning to $n_k$. 

In conclusion, we have found that the quantum kinetic approach has many merits. The phase-space functions have rather intuitive interpretations in terms of distribution functions, and the equations describing how they evolve are simple to solve, at least in the homogeneous limit. Studying the solutions, we were also able to quickly arrive at a precise interpretation of a key
particle number, showing that the quantum kinetic formalism can help clarify certain definitions in a more direct way than other approaches. Due to the generality of the quantum kinetic approach, our considerations can also be extended systematically to
include spatial dependencies and backreaction. Thus, this framework provides a promising path to study the production of particles, such as those possibly constituting dark matter, in complex scenarios while still being close to physical interpretations

\begin{acknowledgments}
The author would like to thank Gert Brodin, Greger Torgrimsson, and Michael Bradley for helpful discussions. 
\end{acknowledgments}


\bibliography{KHWCS}

\end{document}